# Probing Femtosecond Lattice Displacement upon Photo-carrier generation in Lead Halide Perovskite

Giovanni Batignani[1,±], Giuseppe Fumero[1,2,±], Ajay Ram Srimath Kandada[3,±], Giulio Cerullo[4], Marina Gandini[3,4], Carino Ferrante[1], Annamaria Petrozza[3,*], Tullio Scopigno[1,5,*]

[1] Dipartimento di Fisica, Universitá di Roma "La Sapienza", Roma I-00185, Italy

[2] Dipartimento di Scienze di Base e Applicate per l'Ingegneria, Universitá di Roma "La Sapienza", Roma I-00185, Italy

[3] Center for Nano Science and Technology @Polimi, Istituto Italiano di Tecnologia, via Giovanni Pascoli 70/3, 20133, Milan, Italy.

[4] Dipartimento di Fisica, Politecnico di Milano, Piazza L. da Vinci, 32, 20133 Milano, Italy

[5] Istituto Italiano di Tecnologia, Center for Life Nano Science @Sapienza, Roma I-00161, Italy

[±] *Equally contributing authors*

[*] *corresponding authors: tullio.scopigno@roma1.infn.it, annamaria.petrozza@iit.it*

*Electronic properties and lattice vibrations are supposed to be strongly correlated in metal-halide perovskites, due to the "soft" fluctuating nature of their crystal lattice. Thus, unveiling electron-phonon coupling dynamics upon ultra-fast photoexcitation is necessary for understanding the optoelectronic behaviour of the semiconductor. Here, we use impulsive vibrational spectroscopy to reveal ground and excited state vibrational modes of methylammonium lead-bromide perovskite. We observe a coherent phonon at 106 cm$^{-1}$ (13 meV), pertaining to the inorganic octahedral, which is peculiar of the electronic excited state and generated via displacive excitation mechanism. This indicates the formation of a new geometry, reached after a quarter of the phonon period T/4=80 fs, and fully equilibrated within the phonon lifetime of about 1 ps. Our observations unambiguously prove that this mode drives the crystalline distortion occurring upon carrier generation, implying the presence of polaronic effects.*

Solution processed hybrid lead-halide perovskites are an emergent class of materials for efficient optoelectronic technologies[1]. Despite the technological appeal, a comprehensive understanding of their photo-excitation dynamics is still lacking. A crucial, missing information is the nature of carriers, a key issue for the description of transport and recombination processes, two unique characteristics of these materials also in focus of a lively debate[2–4]. In fact, the reported carrier recombination rates[5] are remarkably low, i.e. comparable to the best ones reported for single crystalline semiconductors[2] and orders of magnitude lower than those predicted by the Langevin model[4]. On the other hand, modest charge mobilities have been reported, i.e. comparable to those found in organic (disordered) semiconductors[6]. Given the polar nature of the perovskite lattice, it has been suggested[6], in order to justify the low mobilities, that carriers localize as large polarons in contrast to a pure band like picture where carriers act as delocalized Bloch waves. This picture would be consistent with carrier transport and temperature dependence of the homogenous linewidths of electronic transitions[7], pointing to the presence of strong electron-phonon scattering mechanism. Recently, it has been advanced[8] that photogenerated electrons relax into a distinct dark electronic state which extends the charge carrier lifetime. While the proposed mechanism is compatible with both the band-like and large polaron pictures, an experimental method able to disentangle these effects is still missing. Therefore, both the nature of such state and the reaction pathway which would lead to its population remain unclear[9].

Here, we use a time domain Raman technique, named impulsive vibrational spectroscopy (IVS), to unveil the phonon spectra of the ground and excited electronic states in methylammonium lead bromide perovskite (MAPbBr$_3$) polycrystalline thin films. Raman spectroscopy is an ideal tool to investigate electron-phonon coupling in materials. While ground state vibrational features in lead halide perovskites have been thoroughly addressed[10–13], identifying the correlation between the phonon modes and electronic excitations is one of the major challenges in the field. In order to address this issue, it would be imperative to perform the measurements with a resonant excitation. Detecting spontaneous Raman signals from excited electronic states using a single resonant excitation is not trivial. Moreover, the presence of a large background signal from photoluminescence (PL) may obscure the low frequency region that contains the most relevant modes of the inorganic moiety. Within this context, IVS[14–16] represents a powerful technique to circumvent these limitations and resolve both ground and excited state Raman bands with high spectral resolution. In fact, since the measurement is performed in the time domain, there are no spectral limitations neither artifacts arising from elastic pump-scattering and PL.

In detail, IVS exploits two femtosecond laser pulses to obtain the Raman vibrations of the system under investigation (Figure 1.a). First, Raman interactions convert an incoming optical pump pulse into an impulsive force acting on the solid-state lattice. This results in collective lattice displacements, namely coherent phonons, provided that the pump pulse duration is shorter than the Raman active vibrational period. Then, the interaction with a delayed probe pulse leads to the generation of a third order electronic polarization in the material, which modulates the transmission of the probe. The overall result is the appearance of a time oscillating transmissivity change at the frequencies of the stimulated coherent phonon modes, detected by the time-delayed probe pulse. This oscillating response is always superimposed on a transient absorption (TA) signal, which arises when carriers are photo-generated. The underlying exponential electronic kinetics responsible for such TA background is then subtracted from the detected signal to extract the oscillating temporal dynamics[15,17]. These time-domain signals are then Fourier transformed to obtain the vibrational Raman spectra (see SI for experimental details). The basics of the IVS experiment, performed at the Femtoscopy labs ("Sapienza" University of Rome), are provided in Figure 1.b.

Figure 1.c shows the absorption spectrum of the polycrystalline thin film of MAPbBr$_3$ (see SI for details of sample preparation) where the red and green arrows indicate the 1.86 eV (665 nm) off-resonant and 2.46 eV (503 nm) resonant excitation pulses used in our experiments. The pump pulse duration was 30 (50) fs for the 1.86 (2.46) eV respectively, which is much shorter than the periods of the vibrational modes of interest. Figure 2.a reports the IVS spectra under such excitations. Since the detection of the probe is spectrally resolved, it provides us with a probe-wavelength resolved vibrational coherence map[18,19]. For each excitation energy, Fourier analysis is performed on a specific spectral region of the transmitted white light continuum probe to exclude elastic scattering of the pump. The exact procedure used for the data analysis is detailed in Section 2 in SI. Briefly, for the off-resonance excitation, we consider probe wavelengths from 480 to 600 nm (photon energies from 2.58 to 2.07 eV), while for the resonant excitation we consider the region from

550 to 630 nm (photon energies from 2.25 to 1.97 eV). For both the actinic excitation energies, we observe an underlying transient absorption signal, in agreement with previous reports[20]. In order to factor out cross phase modulation artefacts, due to pump and probe temporal overlap, time traces have only been considered from 300 fs on-wards in the data analysis. After removing the superimposed dynamics from the TA signal, we apply Fourier transformation at each probe wavelength. Zero padding algorithm and Kaiser-Bessel windowing are exploited to enhance the spectral definition (see section 2 of SI). At the end, we obtain 2D vibrational maps shown in Figure 2.a (top and bottom panels). Slices of these maps for selected probe wavelengths are reported in the middle panel, where the green and red shaded spectra correspond to resonant and off-resonant excitations respectively. The obtained Raman spectra are fitted with Lorentzian functions to extract the peak positions of the various modes, which are plotted as a function of probe wavelength in Figure 2.b, for both the pumps. Statistically relevant features are observable in all the probe spectral regions.

In the off-resonant case, we observe predominant peaks at 78 cm$^{-1}$, 90 cm$^{-1}$, 121 cm$^{-1}$, 134 cm$^{-1}$ and 317 cm$^{-1}$. These frequencies are in fair agreement with theoretical calculations and continuous wave (CW) Raman spectra (for which the peak positions of the modes are reported in Fig. 2b as dashed vertical lines [13]). The corresponding amplitudes tend to vanish approaching the absorption maximum, as expected for the ground state IVS scattering mechanism[15]. The low energy modes (< 200 cm$^{-1}$) represent the vibrational degrees of freedom exhibited by the lead-halide octahedral, pointing to the inherent softness of the inorganic cage[10]. The modes which we observe at 90 cm$^{-1}$ and 134 cm$^{-1}$ have been assigned to the symmetric stretch modes of the Br-Pb-Br bond, the mode at 121 cm$^{-1}$ arises out of the asymmetric stretch. Notably, compared to CW Raman, the 90 cm$^{-1}$ mode is slightly red shifted (see Figure 3) and the expected 66 cm$^{-1}$ mode is absent. The mode at 317 cm$^{-1}$ has been previously attributed to the torsional modes of the methyl ammonium cation[10,21,22]. The measured full widths at half maximum ($\Delta\nu_{FWHM}$) of the modes indicate phonon dephasing times[23] $\Gamma^{-1} \approx$ 1 ps.

Notably, in striking contrast with the off-resonant pumping regime, under resonance condition we observe only two dominant modes with constant frequencies, at 106 cm$^{-1}$ and at 121 cm$^{-1}$, all over the probe region (Figure 2). In the case of non-resonant excitation, the coherent phonons are generated exclusively by an impulsive stimulated Raman scattering (ISRS)[14] process, when the pump pulse duration is shorter than the phonon period. Within such a scenario, coherent lattice motions are induced in the electronic ground state and the spectrum obtained must resemble the CW Raman spectrum. Under resonant conditions, the ISRS process undergoes selective resonance enhancement for those ground state vibrational modes (see Figure 2.c) which are coupled to the electronic transition[24,25] via electron-phonon interactions. Accordingly, the 121 cm$^{-1}$ line, which we observe in resonant-IVS, identifies the only ground-state phonon which is strongly coupled with the photo-excited carriers. The energy of this mode, at approximately 15 meV, matches well with the phonon mode identified via temperature dependent photoluminescence spectroscopy[7] and via mid infrared spectroscopy[26]. If the potential energy surface of the excited state is shifted in the coordinate space with respect to that of the ground state, the generation of coherent vibrational modes can also occur via a displacive excitation process (DECP, see Figure 2.d)[27]. In molecular systems, this implies a geometrical rearrangement of the atoms, driven by the dynamics of the vibrational wavepacket in the electronically excited state. In crystalline materials, the same mechanism results in polaronic distortions, namely electron–phonon interactions driving carrier induced shifts of atoms' equilibrium positions and vibration frequencies[28]. Hence, we suggest the 106 cm$^{-1}$ line, appearing only in resonant-IVS, to be DECP generated. Analysis of relative phase (presented in SI) between the observed modes further substantiates this assignment. The relative phase difference between those modes observed in the non-resonant case is approximately null (Figure S1), due to a wavepacket generation occurring in the energy minimum of the stimulated vibrational coordinate (Figure 2.c). On the other hand, there is a non-zero phase difference between the modes at 121 cm$^{-1}$ and 106 cm$^{-1}$ observed in the resonant IVS (Figure S2). This is consistent with the DECP nature of the latter, which originates from the vertical projection of the initial ground state wavepacket far from the equilibrium position of the new lattice geometry (Figure 2.d). Accordingly, this is the mode that drives the crystalline distortion through transient lattice oscillation, leading to a new configuration reached after a quarter of the phonon period T/4= 80 fs, and completely relaxed within the phonon lifetime ($\Gamma^{-1} \approx 1$ ps). Notably, on the timescale of the lattice rearrangement the transient absorption spectrum presents a photo-bleaching (PB) band arising from the state-filling convolved with a derivative like line-shape (see Figure S3 of the SI), which is generally the signature of Coulomb effects on the electronic/excitonic transition[29]. While the PB follows the charge carrier dynamics, living for hundreds of picoseconds[20], the derivative feature decays within the first picosecond, i.e. during the time interval taken by the new lattice geometry to equilibrate, supporting the presence of ionic displacements.

In conclusion, by contrasting the response obtained upon resonant and non-resonant pumping, we demonstrated how IVS can be employed to isolate those phonons coupled to a specific electronic transition in hybrid perovskites. Most importantly, we revealed the key phonon mode, generated via displacive excitation mechanism, which provides the pathway to the photo-induced lattice modification occurring upon carrier generation. Since delocalized carriers in conduction/valence bands would not show an associated pattern of displaced atomic equilibrium positions, our results provide evidence for the polaronic nature of charges in this class of systems. Moreover, they also offer a rationale for the extended carrier lifetime in lead halide perovskites. In order for the carriers to relax back into the ground state, indeed, the lattice would need an additional amount of energy to re-configure itself, which makes such a transition less favorable.

## References

1. Stoumpos, C. C. & Kanatzidis, M. G. The Renaissance of Halide Perovskites and Their Evolution as Emerging Semiconductors. *Acc. Chem. Res.* **48,** 2791–2802 (2015).

2. Zhu, X. Y. & Podzorov, V. Charge Carriers in Hybrid Organic-Inorganic Lead Halide Perovskites Might Be Protected as Large Polarons. *J. Phys. Chem. Lett.* **6,** 4758–4761 (2015).

3. Stranks, S.; Snaith, H.J. Metal-halide perovskites for photovoltaic and light-emitting devices. *Nat. Nanotechnol.* **10,** 391–402 (2015).

4. Herz, L. M. Charge-Carrier Dynamics in Organic-Inorganic Metal Halide Perovskites. *Annu. Rev. Phys. Chem.* **67,** 3.1-3.25 (2016).

5. Srimath Kandada, A. R. & Petrozza, A. Research Update : Luminescence in lead halide perovskites Research Update : Luminescence in lead halide perovskites. *APL Mater.* **91506,** (2016).

6. Yi, H. T., Wu, X., Zhu, X. & Podzorov, V. Intrinsic Charge Transport across Phase Transitions in Hybrid Organo-Inorganic Perovskites. *Adv. Mater.* **28,** 6509–6514 (2016).

7. Wright, A. D. *et al.* Electron–phonon coupling in hybrid lead halide perovskites. *Nat. Commun.* **7,** 11755 (2016).

8. Hutter, E. M. *et al.* Direct–indirect character of the bandgap in methylammonium lead iodide perovskite. *Nat. Mater.* **16,** 115–120 (2017).

9. Moser, J.-E. Perovskite photovoltaics: Slow recombination unveiled. *Nat. Mater.* **16,** 4–6 (2017).

10. Quarti, C., Grancini, G. & Mosconi, E. The Raman Spectrum of the CH3NH3PbI3 Hybrid Perovskite: Interplay of Theory and Experiment. *J. Phys. Chem. Lett.* **5,** 279–284 (2014).

11. Park, B. *et al.* Resonance Raman and Excitation Energy Dependent Charge Transfer Mechanism in Halide Substituted Hybrid Perovskite Solar Cells. *ACS Nano* **9,** 2088–2101 (2015).

12. Brivio, F. *et al.* Lattice dynamics and vibrational spectra of the orthorhombic, tetragonal, and cubic phases of methylammonium lead iodide. *Phys. Rev. B* **92,** 144308 (2015).

13. Leguy, A. M. A. *et al.* Dynamic disorder , phonon lifetimes , and the assignment of modes to the vibrational spectra of methylammonium lead halide perovskites. *Phys. Chem. Chem. Phys.* **18,** 27051–270566 (2016).

14. Dhar, L., Rogers, J. A. & Nelson, K. A. Time-resolved vibrational spectroscopy in the impulsive limit. *Chem. Rev.* **94,** 157–193 (1994).

15. Liebel, M., Schnedermann, C., Wende, T. & Kukura, P. Principles and Applications of Broadband Impulsive Vibrational Spectroscopy. *J. Phys. Chem. A* **119,** 9506–9517 (2015).

16. Kahan, A., Nahmias, O., Friedman, N., Sheves, M. & Ruhman, S. Following photoinduced dynamics in bacteriorhodopsin with 7-fs impulsive vibrational spectroscopy. *J. Am. Chem. Soc.* **129,** 537–546 (2007).

17. Schnedermann, C. *et al.* Vibronic Dynamics of the Ultrafast all-trans to 13-cis Photoisomerization of Retinal in Channelrhodopsin-1. *J. Am. Chem. Soc.* **138,** 4757–4762 (2016).

18. Liebel, M. & Kukura, P. Broad-band impulsive vibrational spectroscopy of excited electronic states in the time domain. *J. Phys. Chem. Lett.* **4,** 1358–1364 (2013).

19. Monacelli, L. *et al.* Manipulating Impulsive Stimulated Raman Spectroscopy with a Chirped Probe Pulse. *J. Phys. Chem. Lett.* **8,** 966–974 (2017).

20. Grancini, G. *et al.* Role of microstructure in the electron–hole interaction of hybrid lead halide perovskites. *Nat. Photonics* **9,** 695–701 (2015).

21. Niemann, R. G. *et al.* Halogen Effects on Ordering and Bonding of CH3NH3+ in CH3NH3PbX3 (X = Cl, Br, I) Hybrid Perovskites: A Vibrational Spectroscopic Study. *J. Phys. Chem. C* **120,** 2509–2519 (2016).

22. Yaffe, O. *et al.* Local Polar Fluctuations in Lead Halide Perovskite Crystals. *Phys. Rev. Lett.* **118,** 136001 (2017).

23. Demtröder, W. *Laser Spectroscopy: Basic Concepts and Instrumentation.* (Springer).

24. Trommer, R. & Cardona, M. Resonant Raman scattering in GaAs. *Phys. Rev. B* **17,** 1865–1876 (1978).

25. Stevens, T., Kuhl, J. & Merlin, R. Coherent phonon generation and the two stimulated Raman tensors. *Phys. Rev. B* **65,** 144304 (2002).

26. Sendner, M. *et al.* Optical Phonons in Methylammonium Lead Halide Perovskites and Implications for Charge Transport. *Mater. Horizons* **3,** 613–620 (2016).

27. Zeiger, H. J. *et al.* Theory for displacive excitation of coherent phonons. *Phys. Rev. B* **45,**

768 (1992).

28. Emin, D. *Polarons*. (Cambridge University Press, 2013).

29. Schmitt-Rink, S., Chemla, D. & Miller, D. Theory of transient excitonic optical nonlinearities in semiconductor quantum-well structures. *Phys. Rev. B* **32,** 6601–6609 (1985).

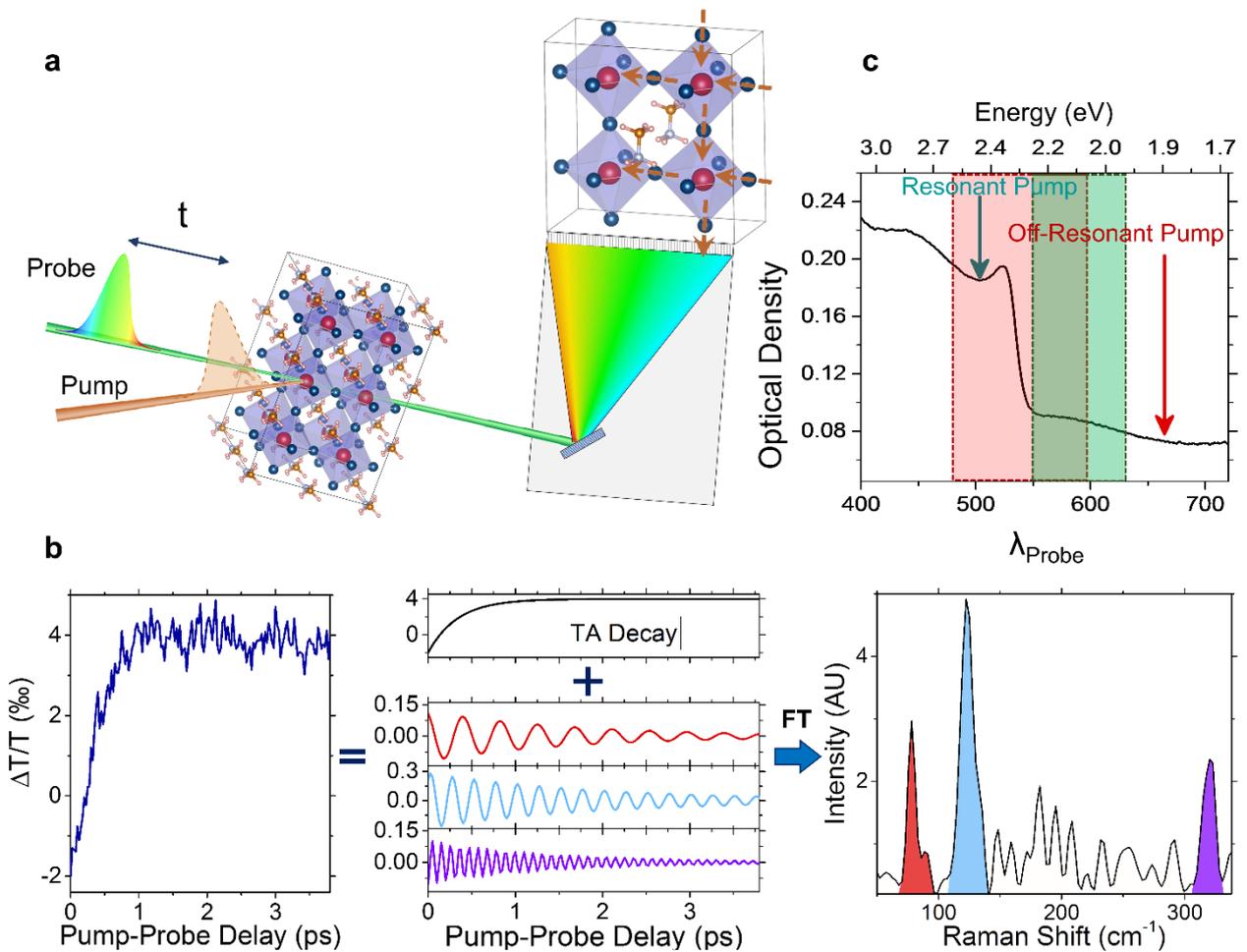

Figure 1: Concept of the Impulsive vibrational spectroscopy experiment on MAPbBr$_3$. (a) IVS pulse scheme and MAPbBr$_3$ crystal structure. After a tunable delay from the interaction with a femtosecond pump pulse, an ultrashort broadband probe pulse interrogates the system and reveals the stimulated lattice vibrations. (b) The experimentally detected differential signal (Left Panel) shows the photo-induced modifications of the transmission profile as a function of the time delay between the two pulses. The signal consists in oscillating components, which carry the phonon frequencies, superimposed to the transient absorption exponential dynamics (Central Panel). The vibrational information is directly obtained by Fourier Transforming the experimental data after the subtraction of the TA decay (Right Panel). (c) Visible absorption spectrum of a thin MAPbBr$_3$ film: the red and green arrows indicate respectively the 1.86 eV off-resonant and the resonant 2.46 eV pump photon energies used in our experiments. The coloured boxes represent the analysed probe spectral regions, under the corresponding excitation regime.

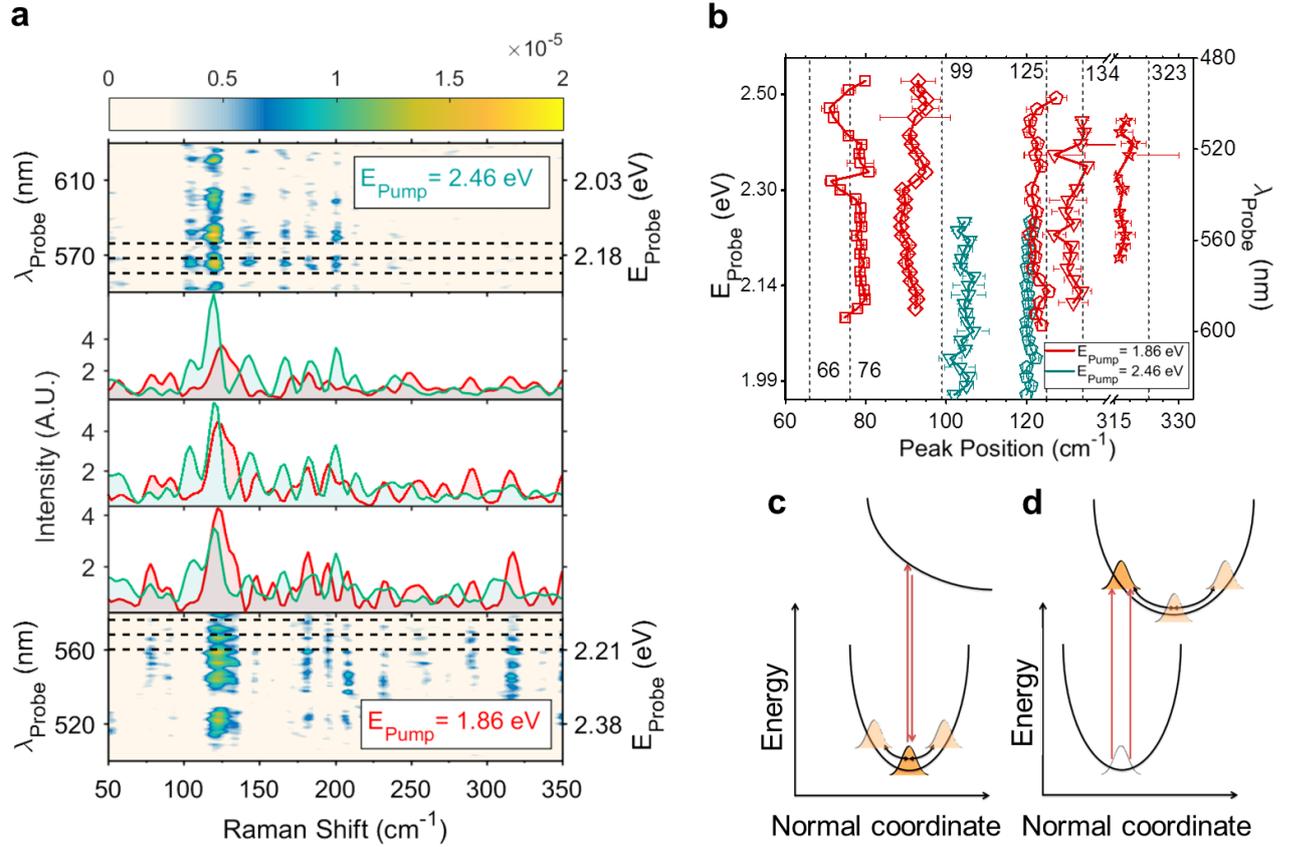

Figure 2: Impulsive vibrational spectroscopy measurements of MAPbBr$_3$ under different resonance regimes. (a) Top and bottom panels report the probe wavelength resolved IVS maps, upon resonant (E$_{Pump}$ = 2.46 eV) and non-resonant (E$_{Pump}$ = 1.86 eV) excitations, respectively. The maps have been obtained by Fourier transforming the oscillating component of the transient absorption data (see Fig. 1), to retrieve the vibrational spectra. The three 1D plots in the central panels represent slices of the maps averaged within a 5 nm region from selected wavelengths (highlighted by the black dashed lines in the maps). Green and red shaded Raman spectra refer to resonant and non-resonant conditions, respectively. (b) Fitted peaks positions of the measured Raman modes as a function of the probed wavelength. Statistically relevant features show a constant peak position in all the probe spectral regions. Red and green symbols refer to Raman modes obtained by pumping at 1.86 eV and 2.46 eV, respectively. Horizontal dotted lines indicate the position of MAPbBr$_3$ ground state vibrational modes reported in literature[21]. (c-d) Representations of the ISRS and DECP processes, respectively. The red arrows indicate the double interactions with the pump pulse, which generates a vibrational coherence in the ground and excited state[18]. In ISRS, immediately after photoexcitation, the vibrational wavepacket is peaked at the equilibrium position and then starts oscillating along the normal mode coordinate. In DECP, the ground state wavepacket is projected onto the excited state, where it begins oscillating from a starting position far from the equilibrium.